\documentclass[pdflatex,sn-nature]{sn-jnl}%

\usepackage{graphicx}%
\usepackage{multirow}%
\usepackage{amsmath,amssymb,amsfonts}%
\usepackage{amsthm}%
\usepackage{mathrsfs}%
\usepackage[title]{appendix}%
\usepackage{xcolor}%
\usepackage{textcomp}%
\usepackage{manyfoot}%
\usepackage{booktabs}%
\usepackage{algorithm}%
\usepackage{algorithmicx}%
\usepackage{algpseudocode}%
\usepackage{listings}%
 \usepackage{gensymb}%
 \usepackage{type1ec}%

\theoremstyle{thmstyleone}%

\theoremstyle{thmstyletwo}%

\theoremstyle{thmstylethree}%

\begin{document}

\title[Emergent domain topology in the multiferroic hexagonal manganites]{Emergent domain topology in the multiferroic hexagonal manganites}

\author*[1]{\fnm{Aaron Merlin} \sur{M\"{u}ller}}\email{aaron.mueller@mat.ethz.ch}

\author[1,2]{\fnm{Lukas} \sur{ Heckendorn}}

\author[1]{\fnm{Manfred} \sur{Fiebig}}

\author[1]{\fnm{Thomas} \sur{Lottermoser}}

\affil[1]{\orgdiv{Department of Materials}, \orgname{ETH Zurich}, \orgaddress{\street{Vladimir-Prelog-Weg 4}, \city{Zurich}, \postcode{8093}, \country{Switzerland}}}

\affil[2]{Current address:  \orgdiv{Institute for Quantum Electronics}, \orgname{ETH Zurich}, \orgaddress{\city{Zurich}, \postcode{8093}, \country{Switzerland}}}

\abstract{Emergent topological phenomena in multiferroic materials arise from the intricate coupling between structural, electric, and magnetic order parameters. Hexagonal manganites provide a paradigmatic platform for such studies. These compounds exhibit a strongly coupled distortive-improper ferroelectric order, arising from trimerizing lattice distortions, and a 120° noncollinear antiferromagnetic spin structure. While their two-dimensional domain topology has been extensively studied, the full three-dimensional multiferroic domain architecture has remained largely unexplored, mainly due to the experimental challenges of probing bulk structures beyond surfaces. Here, we employ a Landau free-energy framework combined with large-scale phase-field simulations to reveal the intricate three-dimensional multiferroic domain network of hexagonal manganites. We demonstrate that the coupling between the structural and antiferromagnetic order parameters gives rise to a rich variety of three-dimensional topological features. In particular, these features give rise to an attraction between different types of domain walls. Moreover, we identify bifurcations of vortex-like lines at domain-wall intersections, a phenomenon that can exist only in three dimensions and fundamentally alters the topology of the domain network. Our results provide a comprehensive theoretical basis for understanding three-dimensional domain interactions in multiferroics and highlight the essential role of dimensionality in coupling improper ferroelectricity, magnetism, and topology in hexagonal manganites.}

\maketitle

\section{Introduction}

Magnetoelectric multiferroics, that is, systems exhibiting coupled magnetic and ferroelectric order, are a promising class of materials for a new generation of devices making use of their spin-charge coupling in the form of ultra-low-power energy consumption \cite{trassin_low_2015, spaldinAdvancesMagnetoelectricMultiferroics2019}, magnetoelectric storage \cite{allibeRoomTemperatureElectrical2012, heron_deterministic_2014}, radio and microwave applications \cite{sunVoltageControlMagnetism2012, linIntegratedMagneticsMultiferroics2016}, and opto-spintronics \cite{nemecAntiferromagneticOptospintronics2018}. This prospective use depends on the distribution and control of the 3D magnetic and electric domains and their interaction and underlines the importance of a comprehensive understanding of domain patterns in multiferroics.

In both experiment and theory, investigation of ferroic domain patterns is often limited to two dimensions, represented by the surface of or a cross section through a 3D sample, or by the effectively 2D domain patterns in thin films. Probing 3D textures experimentally is challenging, 
and applications of contemporary 3D topography methods are limited \cite{kampfeOpticalThreedimensionalProfiling2014a, donnellyThreedimensionalMagnetizationStructures2017}. On the theory side, simulations of 3D domain patterns are plagued by high computational cost and impediments in visualization and data analysis. 
However, changing the dimensionality of a system from two to three dimensions might result in significant qualitative changes in domain patterns and novel topological configurations that are absent in 2D \cite{selingerIntroductionTopologicalDefects2024}. 
Their exploration would also be relevant for technological applications, where a true understanding of 3D order is essential for storing significantly more information than in 2D systems.

\begin{figure}[ht!]
    \centering
    \includegraphics[width=0.4\linewidth]{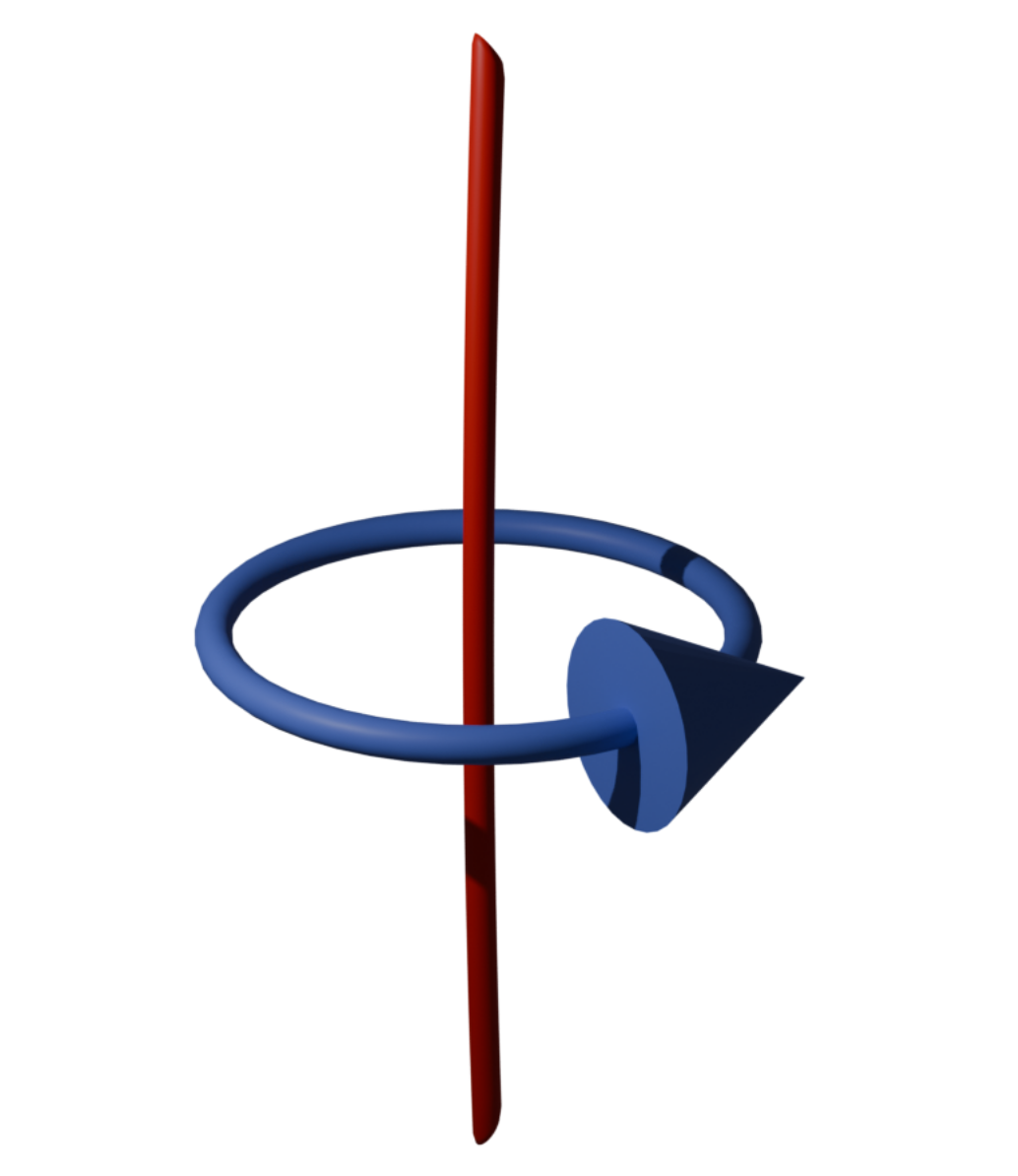}
    \caption{$\mid$ \textbf{Visualization of a vortex line.} A vortex line corresponds to a 1D line in 3D space. Following a path that leads around this line, the angular coordinate or complex phase of the order parameter performs a full 360° rotation or multiples thereof.}
    \label{fig:Vortices_and_pseudo-vortices}
\end{figure}

Topological defects can appear in 3D domain configurations of order parameters exhibiting a periodic angular coordinate or complex phase. Following a path around such a defect, as illustrated in Fig.\ \ref{fig:Vortices_and_pseudo-vortices}, one obtains a full 360° rotation of the angular coordinate or complex phase (or multiples thereof); such defects are referred to as one-dimensional vortex lines. These vortex lines cannot be destroyed under small perturbations of the 3D domain distributions. Hence, the vortex lines are topologically protected
\cite{selingerIntroductionTopologicalDefects2024}.

One of the most extensively studied classes of multiferroic materials are hexagonal manganites \textit{R}MnO$_3$, with \textit{R} = Sc, Y, In, Dy-Lu \cite{sim_hexagonal_2016}. 
A unit-cell-trimerizing tilt of MnO$_5$ bipyramids emerges below 1200-1700\ K as the primary order of the system,  
driving the emergence of an improper electric polarization as a secondary order.  
Additional antiferromagnetic 120° Mn$^{3+}$ spin order emerges below 65-130\ K \cite{fiebigSpinrotationPhenomenaMagnetic2003}.  

Among these, the structural order exhibits topological defects in its domain pattern, which take the form of 1D vortex lines in a 3D system, as introduced above. If one considers a 2D surface intersecting with a vortex line, the vortex line will become a point in the form of a vortex or antivortex. Both, structural vortex lines and vortex-antivortex pairs, have been extensively investigated in recent work \cite{linTopologicalDefectsRelics2014,spaldinMultiferroicsCosmicallyLarge2017,holtzTopologicalDefectsHexagonal2017a,yangElectricFieldDriven2017,yangSpatialAnisotropyTopological2017,xue_topological_2018, mokhtarThreedimensionalDomainIdentification2024,najeebThreedimensionalImagingTopologically2025}.

In contrast, the domain pattern of the antiferromagnetic order of the hexagonal manganites has, to date, not been extensively probed in three dimensions. Whereas the antiferromagnetic order at bulk surfaces is well-understood from experimental 2D studies \cite{fiebigObservationCoupledMagnetic2002, chaeDirectObservationProliferation2012, gengCollectiveMagnetismMultiferroic2012, giraldo_magnetoelectric_2021}, extending these investigations into the third dimension has proven challenging due to a lack of spatially resolving 3D magnetic imaging techniques   
\cite{giraldo_magnetoelectric_2021, kampfeOpticalThreedimensionalProfiling2014a}.

In this work, we present a comprehensive theoretical description of the 3D multiferroic domain structure of hexagonal manganites. 
We visualize the antiferromagnetic and distortive-ferroelectric domains and their topological features and reveal a bifurcation of pseudo-vortex lines at intersections of magnetic and electric domain walls, 
a topological phenomenon that can only exist in three dimensions. 
Starting from the Landau free-energy expansion \cite{artyukhin_landau_2014, giraldo_magnetoelectric_2021} we perform phase-field simulations of the hexagonal manganites in their multiferroic phase and describe how the coupling between the magnetic and the structural orders transfers the topological defects from one to the other. We describe three types of topological features occuring in three dimensions: multiferroic vortex lines, pseudo-vortex lines at domain wall intersections, and bifurcations of pseudo-vortex lines. 
This comprehensive description lays the foundation for 3D experimental studies that could be based on techniques such as X-ray tomography or Bragg coherent diffraction imaging \cite{donnellyThreedimensionalMagnetizationStructures2017, najeebThreedimensionalImagingTopologically2025}.
  
\section{Multiferroic hexagonal manganites} \label{sec:hexman}

\begin{figure}[ht!]
    \centering
    \includegraphics[width=0.65\linewidth]{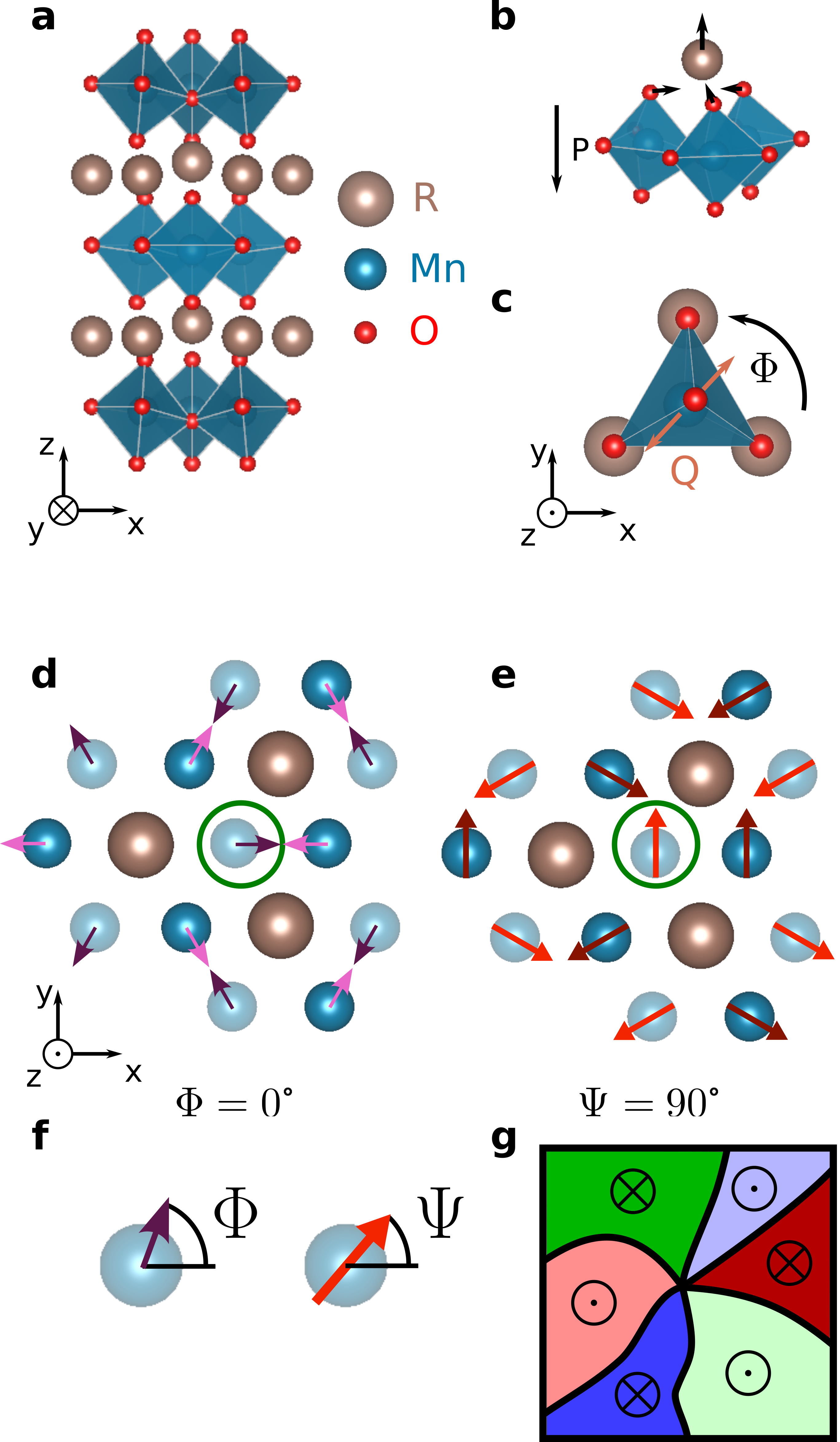}
    \caption{$\mid$ \textbf{Structural and magnetic order of hexagonal manganites.} \textbf{a} Structural order of hexagonal manganites, corresponding to the zone-boundary mode $K_3$. The mode consists of a MnO$_5$ bipyramid tilt and an up-down order of \textit{R}-atoms. \textbf{b} Visualization of the bipyramid tilt and the shift of the \textit{R} atoms. Where all three bipyramids tilt towards a common center, the \textit{R}-atoms shift away from the bipyramids, while the other \textit{R}-atoms move in the opposite direction. \textbf{c} The structural order parameter components $Q$ and $\Phi$ correspond to the amplitude and azimuthal angle of the bipyramid tilt. \textbf{d} Top view of the structural order with arrows illustrating bipyramid tilts with respect to the associated Mn$^{3+}$ ions. \textbf{e} Top view of the antiferromagnetic order of the system on the example of ErMnO$_3$, with arrows corresponding to the spins of the Mn$^{3+}$ ions. In \textbf{d} and \textbf{e}, the reference Mn$^{3+}$ ion is ion marked with a green circle. Different shades of Mn$^{3+}$ ions and arrows correspond to atoms in adjacent Mn-O layers. \textbf{f} Definition of $\Phi$ and $\Psi$ in reference to \textbf{d} and \textbf{e}, respectively. \textbf{g} 2D $z$-cut of a structural domain pattern, with 60° structural domain walls in $\Phi$ and six $\Phi$-domains meeting at a vortex. The secondary ferroelectric polarization follows an alternating pattern, depicted with $\odot$ and $\otimes$ symbols.}
    \label{fig:Intro_Crystal_Structure}
\end{figure}

The crystal structure of the hexagonal manganites 
is visualized in Fig.\ \ref{fig:Intro_Crystal_Structure}a,b. 
The primary order of the system is the distortive-structural one and corresponds to the zone-boundary mode $K_3$. It describes a tilt of sets of three MnO$_5$ bipyramids towards a common center with the \textit{R}-atom shifting away from or toward this center. This primary structural order is then coupled to the secondary ferroelectric order corresponding to the polar mode $\Gamma_2^-$ \cite{fennie_ferroelectric_2005}. We describe (i) the structural mode with an order parameter $\mathbf{Q}$ that corresponds to a tilt of the MnO$_5$ bipyramids, and (ii) the polar mode with order parameter $\mathcal{P}$, which is proportional to the polarization $P$. For $\mathbf{Q}$, it is convenient to describe the distortive-structural order in polar coordinates with magnitude $Q$ and angle $\Phi$, which corresponds to, respectively, the amplitude and the azimuthal angle of the tilt axis of a given bipyramidal tilt in the unit cell, see Fig.\ \ref{fig:Intro_Crystal_Structure}c. As shown in Fig.\ \ref{fig:Intro_Crystal_Structure}d, the orientation of a single bipyramidal tilt defines the tilt of all other bipyramids in the unit cell, so it is sufficient to parametrize the order with a single tilt angle. 

The structural and ferroelectric order of hexagonal manganites can be described with a Landau-theoretical approach developed by Artyukhin et al.\ \cite{artyukhin_landau_2014},
\begin{equation}
    \begin{split}
        f = & \frac{a}{2} Q^2 + \frac{b}{4} Q^4 + \frac{Q^6}{6} (c + c' \text{cos} \, 6 \Phi ) \\
        - &  g Q^3 \mathcal{P} \text{cos} 3 \Phi + \frac{g'}{2} Q^2 \mathcal{P}^2 + \frac{a_\mathcal{P}}{2} \mathcal{P}^2 \\
        + & \frac{1}{2} \sum \limits_{i = x,y,z} s^i_Q (\partial_i Q \partial_i Q + Q^2 \partial_i \Phi \partial_i \Phi) + s_\mathcal{P}^i \partial_i \mathcal{P} \partial_i \mathcal{P},
    \end{split}
    \label{eq:LandauTheory}
\end{equation}
with parameters $a = -2.626 \, \text{eV \AA}^{-2}$, $b = 3.375 \, \text{eV \AA}^{-4}$, $c = 0.117 \text{eV \AA}^{-6}$, $c' = 0.108 \text{eV \AA}^{-6}$, $a_\mathcal{P} = 0.866 \, \text{eV \AA}^{-2}$, $g = 1.945 \, \text{eV \AA}^-4$, $g' = 9.931 \, \text{eV \AA}^{-4}$, $s_Q^z = 15.40 \, \text{eV}$, $s_Q^x = 5.14 \, \text{eV}$, $s_\mathcal{P}^z = 52.7 \, \text{eV}$, and $s_\mathcal{P}^x = 8.88 \, \text{eV}$ in the case of YMnO$_3$, based on density functional theory (DFT) calculations \cite{artyukhin_landau_2014, xue_topological_2018}. The first line of Eq.\ \ref{eq:LandauTheory} describes the primary structural order, whereas the second line describes the coupling between the primary structural and the improper ferroelectric order. The third line contains stiffness terms that describe penalties for domain walls.

The Landau-theory-approach thus contains all ingredients to describe the characteristic domain pattern of the primary structural and the secondary ferroelectric order. The anisotropy term $\text{cos} \, 6 \Phi$ results in six minima in $\Phi$ that consecutively differ by 60°. The minima correspond to six different domain states arranged with continuously increasing or decreasing value of $\Phi$ around a vortex line. The domains are separated by 60° domain walls, as shown in Fig.\ \ref{fig:Intro_Crystal_Structure}g.
The term $g \, Q^3 \, \mathcal{P} \, \text{cos} 3 \Phi$ couples the tilt angle $\Phi$ and the polar mode $\mathcal{P}$ such that a 60° change in $\Phi$ results in a sign change in $\mathcal{P}$. This results in the alternating polarization pattern in Fig.\ \ref{fig:Intro_Crystal_Structure}g. The zero-dimensional vortex in the 2D visualization of Fig.\ \ref{fig:Intro_Crystal_Structure}g extends into a 1D vortex line in a 3D system as discussed before.

We now turn towards the triangular antiferromagnetic Mn$^{3+}$ spin order of the hexagonal manganites. 
Triangular antiferromagnetic order emerges below 65-130\ K \cite{fiebigSpinrotationPhenomenaMagnetic2003}.
The ErMnO$_3$-like spin arrangement according to the B$_2$-representation, as an example of such order, is visualized in Fig.\ \ref{fig:Intro_Crystal_Structure}e. 
As shown in previous experimental and theoretical work \cite{dasBulkMagnetoelectricityHexagonal2014, giraldo_magnetoelectric_2021, tosicInfluenceTriangularMnO2022b}, the antiferromagnetic order strongly couples to the structural-distortive order. We describe the antiferromagnetic order with an order parameter $\Psi$, which corresponds to the angle of the Mn$^{3+}$ spins in the basal plane.  
Just like the structural order, the magnetic order can be parametrized by a single angle $\Psi$, which defines all other spin orientations as shown in Fig.\ \ref{fig:Intro_Crystal_Structure}e. Note, however, that the value of $\Psi$
depends on the $\textit{R}$-compound \cite{tosicInfluenceTriangularMnO2022b}, for which we remain with Er in the following. 

Landau theory describes the free energy of the antiferromagnetic order by
\begin{equation}
    f = A \, Q^2 \, \text{cos}^2 (\Psi - \Phi) + s_\Psi (\nabla \Psi)^2
    \label{eq:Coupling}
\end{equation}
with parameters $s_\Psi = 742 \, \text{meV \AA}^2$ and $A = 2.13 \, \text{meV}$, which are derived from first-principles calculations of ErMnO$_3$ \cite{giraldo_magnetoelectric_2021, tosicInfluenceTriangularMnO2022b}. The first term corresponds to the coupling between the structural and the antiferromagnetic order and the second term is a stiffness term that penalizes magnetic domain walls, in particular those involving large spin-reorientation angles. We must have $s_\Psi > 0$ to ensure stability of the system.

We use phase-field simulations (also known as time-dependent Ginzburg-Landau simulations) to describe the antiferromagnetic order in three dimensions \cite{xue_strain-induced_2017,xue_topological_2018, bortis_manipulation_2022,sandvik_pressure_2023,mullerEffectDepolarizingField2024, zahnReversibleLongrangeDomain2025}. We start by initializing the structural and the ferroelectric order randomly, parametrized by $(Q,\Phi)$ and $\mathcal{P}$ respectively, in the computational domain. The evolution of the system is then given by the time-dependent Ginzburg-Landau equation
\begin{equation}
    \frac{\partial \eta}{\partial t} = - \frac{\delta f }{ \delta \eta},
    \label{eq:Landau_Derivative}
\end{equation}
where $\eta$ is one of the order parameters $\mathbf{Q}$, $\mathcal{P}$, or $\Psi$. The expression $\delta / \delta \eta$ corresponds to a functional derivative. We let the structural and improper ferroelectric orders relax until a domain pattern has formed. We subsequently freeze the evolution of the ferroelectric order and initialize a random magnetic order on top of it. Then the magnetic order is relaxed, once again using the dynamics from Eq.\ \ref{eq:Landau_Derivative}, to obtain an antiferromagnetic domain pattern. 
This two-step procedure is motivated by the fact that the magnetic order has a much lower transition temperature than the ferroelectric order, permitting us to simulate the onset of the magnetic order with an evolution occurring much faster than any remaining changes in the ferroelectric order. 
Computational details are discussed in Appendix\ \ref{app:NumericalMethods}. This procedure allows us to obtain an unprecedented description of the antiferromagnetic order and its magnetoelectric coupling to the structural, respectively ferroelectric order, in three dimensions.

\section{Types of magnetostructural domain walls}

\label{sec:EquationDiscussion}

\begin{figure}[ht!]
    \centering
    \includegraphics[width=1.0\linewidth]{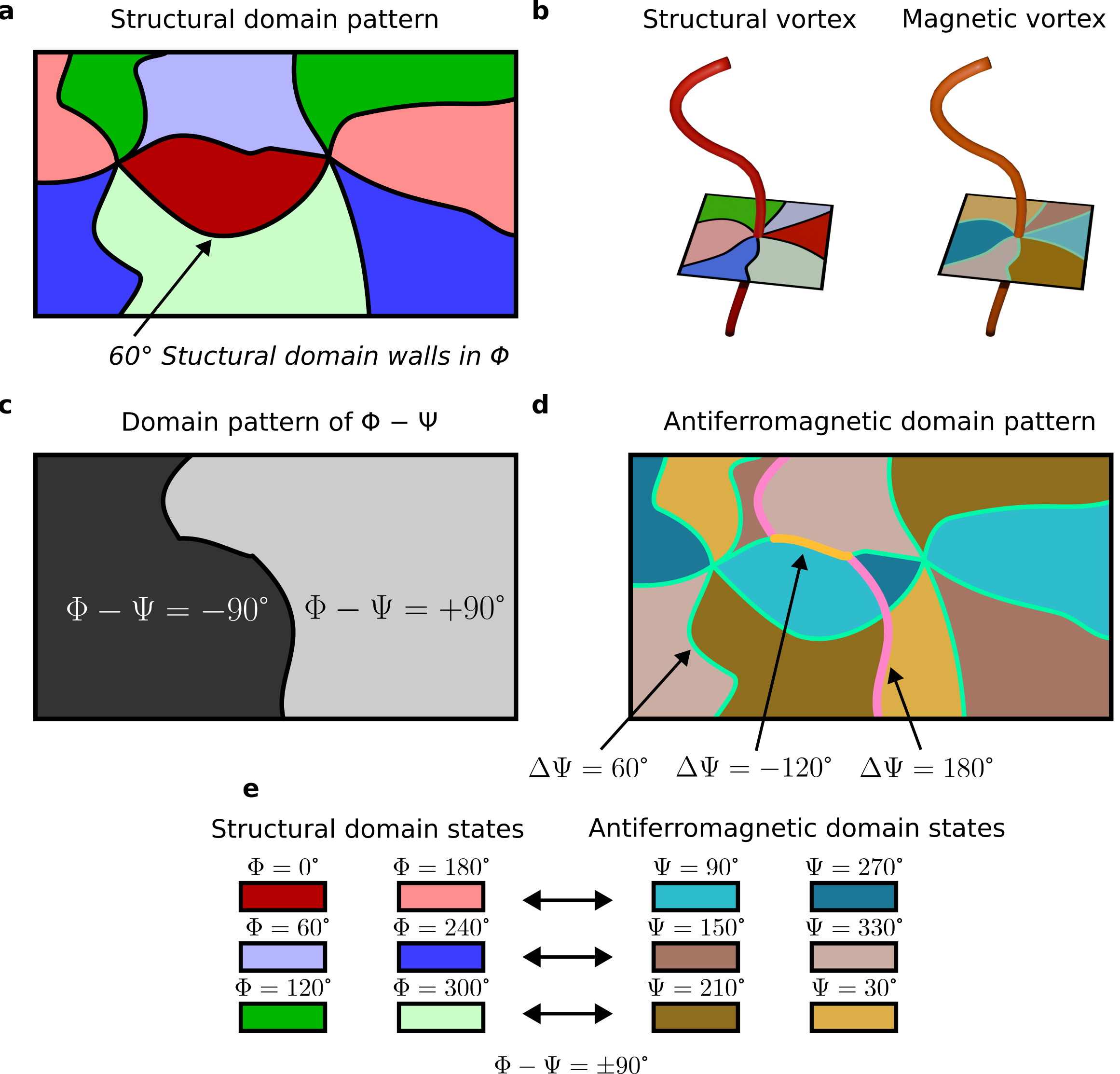}
    \caption{$\mid$ \textbf{Illustration of structural and magnetic domain patterns.} \textbf{a} Schematic domain pattern of the structural order with 60° domain walls in $\Phi$. \textbf{b} Illustrations of structural and magnetic six-fold vortex lines. In minimization of Eq.\ \ref{eq:Condition}, they show a one-to-one correspondence. \textbf{c} Ising-like magnetostructural $\Phi - \Psi$ domain pattern. The two domain states with $\Phi - \Psi = \pm 90$° are separated by a magnetostructural 180° domain wall. \textbf{d} Magnetic domain pattern. Note that the correlation between the domain pattern in \textbf{a} and the $\Phi - \Psi$ domain pattern in \textbf{c} results in three types of magnetic domain walls with a change of $\Psi$ by 60°, 180° or $-$120°. \textbf{e} Legend of magnetic and structural domain states. Pairs of $\Phi$ and $\Psi$ in the same row fulfill the condition $\Phi - \Psi = \pm 90$°.}
    \label{fig:Analytical_Discussions_Of_Domain_Walls}
\end{figure}

Before discussing the results of the simulations, it is insightful to look at the different types of magnetostructural domain walls we expect based on Eq.\ \ref{eq:Coupling}. This will lead to a discussion of the intersection of these walls and their topological properties. We first have the purely structural 60° domain walls in $\Phi$ that are associated with the formation of six-fold distortive and ferroelectric vortex lines as discussed in Section \ref{sec:hexman} and sketched in Fig.~\ref{fig:Analytical_Discussions_Of_Domain_Walls}a. Finding the energy minimum for Eq.\ \ref{eq:Coupling} then leads to the condition 
\begin{equation}
    \Psi - \Phi = \pm 90 \degree,
    \label{eq:Condition}
\end{equation}
which follows from minimization of the first term, $A \, Q^2 \, \text{cos}^2 (\Psi - \Phi)$, under the assumption of $A>0$ (for reason, see below). To obtain the solution in Eq.\ \ref{eq:Condition}, any 60° change in $\Phi$ occurring at a structural domain wall must be accompanied by a simultaneous change in $\Psi$, and hence by an overlapping magnetic domain wall. The structural domain pattern in $\Phi$ is thus transferred to an identical magnetic domain pattern in $\Psi$. Although the first term in Eq.\ \ref{eq:Coupling} still allows an arbitrary choice of the sign of $\Psi-\Phi$, this sign relation is determined by the second term in Eq.\ \ref{eq:Coupling}. It describes the energy introduced by a magnetic domain wall according to $s(\nabla\Psi)^2$.
Due to the term $\nabla\Psi$, the reorientation of the Mn$^{3+}$ spin angle across the wall should be as small as possible, and the smallest value we can have here is $\Delta\Psi=\pm 60$°. In combination with Eq.\ \ref{eq:Condition} this means that any structural domain wall with $\Delta\Phi=\pm 60$° is accompanied by an overlapping magnetic domain wall with $\Delta\Psi=\pm 60$°. In other words, as depicted in Fig.~\ref{fig:Analytical_Discussions_Of_Domain_Walls}b, the structural domain vortex structure in $\Phi$ is copied into an identical magnetic vortex structure in $\Psi$, without sign change in $\Psi-\Phi$ around the vortex. A similar distribution of antiferromagnetic vortices has been observed in a planar $\alpha$-Fe$_2$O$_3$/Co heterostructure \cite{chmielObservationMagneticVortex2018}. There, however, it is induced by interfacial exchange and not, as in our case, by an intrinsic bulk magnetostructural effect.

Keeping the condition of $\Psi-\Phi=\pm 90$° for the first term, which guides the bulk coupling, violation of the minimization condition of the second term, which deviates from zero at the domain walls only, leads to two additional types of magnetic domain walls. First, within a structural domain where $\Phi$ is constant, we may have a change from $\Psi-\Phi=\pm 90$° to $\Psi-\Phi=\mp 90$°, and hence a purely magnetic $180$° domain wall in $\Psi$. The resulting Ising-like domain pattern is sketched in Fig.\ \ref{fig:Analytical_Discussions_Of_Domain_Walls}c. Second, we may have the combination of a structural $60$° domain wall in $\Phi$ with a $180$° magnetic domain wall in $\Psi$. The result is a magnetostructural $-120$° domain wall, where the minus sign indicates that the change of $60$° in $\Phi$ and of $120$° in $\Psi$ occurs in opposite directions \cite{giraldo_magnetoelectric_2021}.

In summary, we have thus identified three types of magnetic domain walls corresponding to a reorientation of the Mn$^{3+}$ spin angle by $60\degree$, $180\degree$, or $- 120\degree$. In our consideration, the choice of $A>0$ in Eq.\ \ref{eq:Condition} was motivated by the value of $A = +2.13 \, \text{meV}$ obtained from first-principles calculations of ErMnO$_3$ \cite{giraldo_magnetoelectric_2021}. This results in the characteristic ErMnO$_3$-like antiferromagnetic order with a Mn$^{3+}$ spin angle of $\pm 90\degree$ as sketched in Fig.~\ref{fig:Analytical_Discussions_Of_Domain_Walls}e. Choosing $A < 0$ instead, one would obtain YMnO$_3$-like antiferromagnetic order order with $\Psi - \Phi = 0\degree, 180\degree$ \cite{artyukhin_landau_2014}. Note that the conclusions made in this work for the ErMnO$_3$-like antiferromagnetic order also apply to the YMnO$_3$-like order, with the only difference being the relative shift in $\Psi$ by $90\degree$.

\subsection{Topological magnetostructural phenomena}

\begin{figure}[ht!]
    \centering
    \includegraphics[width=0.7\linewidth]{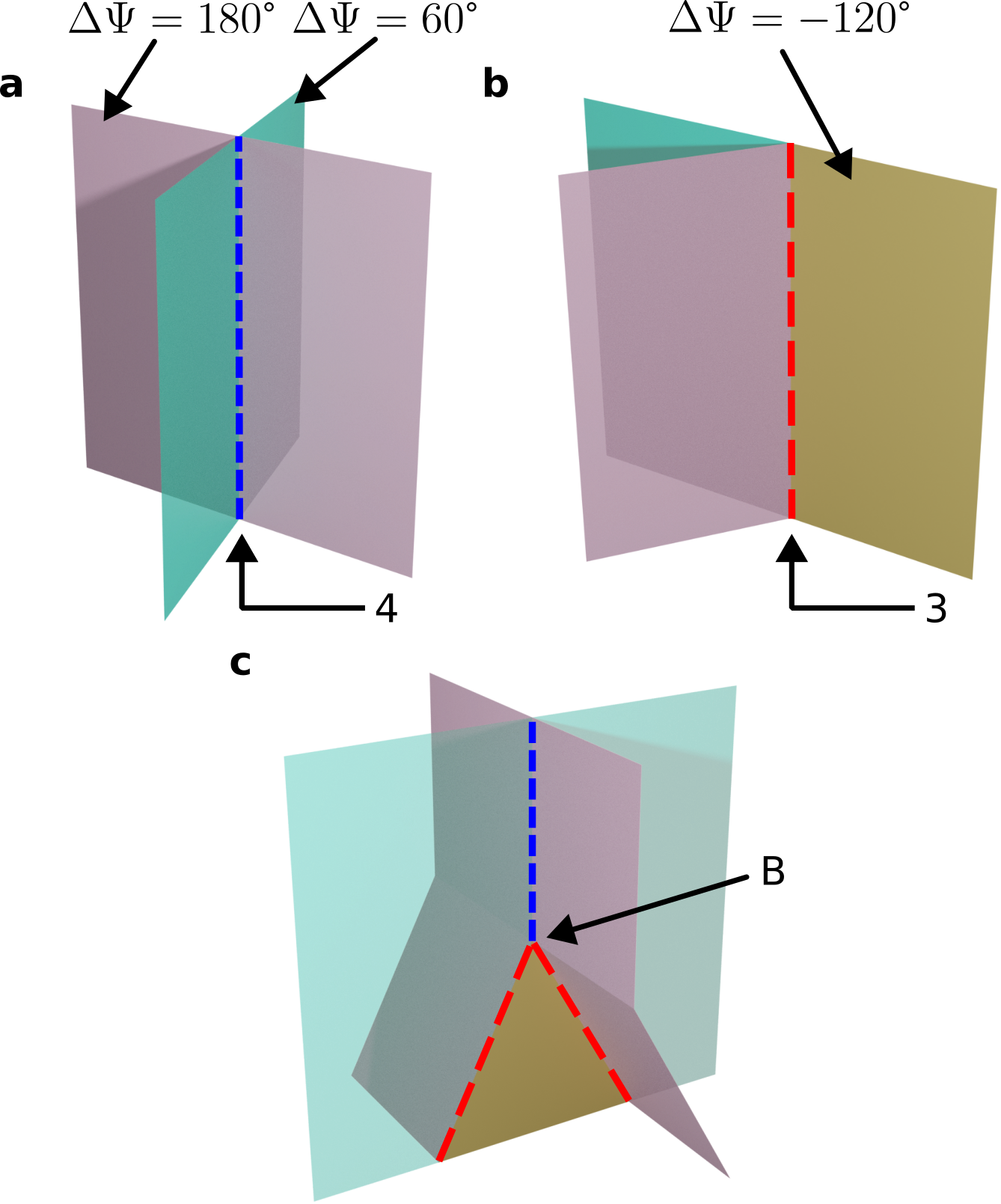}
    \caption{$\mid$ \textbf{Pseudo-vortex lines and bifurcation.} 
    \textbf{a} Four-fold pseudo-vortex line (blue dashed line, marked with `4') as an intersection of a 180° magnetic domain wall (purple) and a 60° magnetic domain wall (turquoise). \textbf{b} Three-fold pseudo-vortex line (red dashed line, marked with `3') as a junction of a 180° and a 60° magnetic domain wall, forming a $-$120° magnetic domain wall (ocher) in the process. \textbf{c} Bifurcation of two three-fold pseudo-vortex lines merging at the point marked `B' to form a four-fold pseudo-vortex line.}
    \label{fig:Topological_Features_3D}
\end{figure}

The intersection of magnetic and structural domain walls results in the emergence of additional topological features. If a 60° and a 180° $\Psi$ domain wall intersect, as shown in Fig.\ \ref{fig:Topological_Features_3D}a, the intersection, shown as a blue dashed line, is also a topologically protected feature. However, its topological protection does not stem from a 360° continuous
rotation of the order parameter but from the fact that domain walls form closed surfaces that cannot end in the bulk of the system. In practice, they behave very similar to the six-fold vortex lines, and hence, we call these features `pseudo-vortex' lines, motivated by previous terminology \cite{giraldo_magnetoelectric_2021}.

In addition to the intersection of a 60° and a 180° magnetic domain wall, which represents a four-fold pseudo-vortex line, we also find a three-fold pseudo-vortex line 
where 60° and 180° domain walls merge into a $-120$° domain wall. This is shown as red-dashed line in Fig.\ \ref{fig:Topological_Features_3D}b.

Finally, a bifurcation point of pseudo-vortex lines, i.e.\ a point where two three-fold pseudo-vortex lines join to form a four-fold pseudo-vortex line, is a topological feature that has no anaologue in two dimensions and is therefore exclusive to the 3D system. It is visualized in Fig.\ \ref{fig:Topological_Features_3D}c. 
As we will see in Sec.\ \ref{sec:pseudo-vortices}, its existence has fundamental consequences for the formation of the network of magnetostructural domains and underlines the need to study the latter in three dimensions for true understanding.

\section{Results and discussion}

Figure \ref{fig:Data_VTK_3D}a gives an overview of the simulated 3D magnetic domain distribution in hexagonal ErMnO$_3$. It shows the 60°, 180° and $-$120° magnetic domain walls in turquoise, purple, and ocher respectively. It also visualizes how the three types of domain walls extend into three dimensions, with a 2D cut at the $z$- and $y$- surfaces showing the topological features discussed in Sec.\ \ref{sec:pseudo-vortices} below. A more comprehensive visualization is given in supplemental movies 1 and 2, where we depict the same data as in Fig.\ \ref{fig:Data_VTK_3D}a using multiple 2D cuts through the domain walls. These movies and additional visualization of the data in slices along all three directions in supplemental data A-C support the a detailed examination of all the topological features in the system that we discuss in detail now.

\subsection{Domain walls and six-fold vortices}

We start by discussing the magnetic domain walls of the system. According to Fig.\ \ref{fig:Data_VTK_3D}a, all three types of $\Psi$ domain walls, that is, 60°, 180°, and $-$120°, exist in the simulation and become 2D manifolds in a 3D system. The figure also shows the numerous intersections between domain walls that we discussed. In Fig.\ \ref{fig:Data_VTK_3D}b, we visualize the six-fold magnetic vortex lines along with a 2D slice as an alternative visualization of the data in Fig.\ \ref{fig:Data_VTK_3D}a. This visualization shows the course of the six-fold vortex lines in the system with further support by supplemental movie 3, where consecutive 2D slices show how the domain walls follow the six-fold vortex lines.

\subsection{Pesudo-vortices and pseudo-vortex bifurcations} 

\label{sec:pseudo-vortices}

\begin{figure}[ht!]
    \centering
    \includegraphics[width=1.\linewidth]{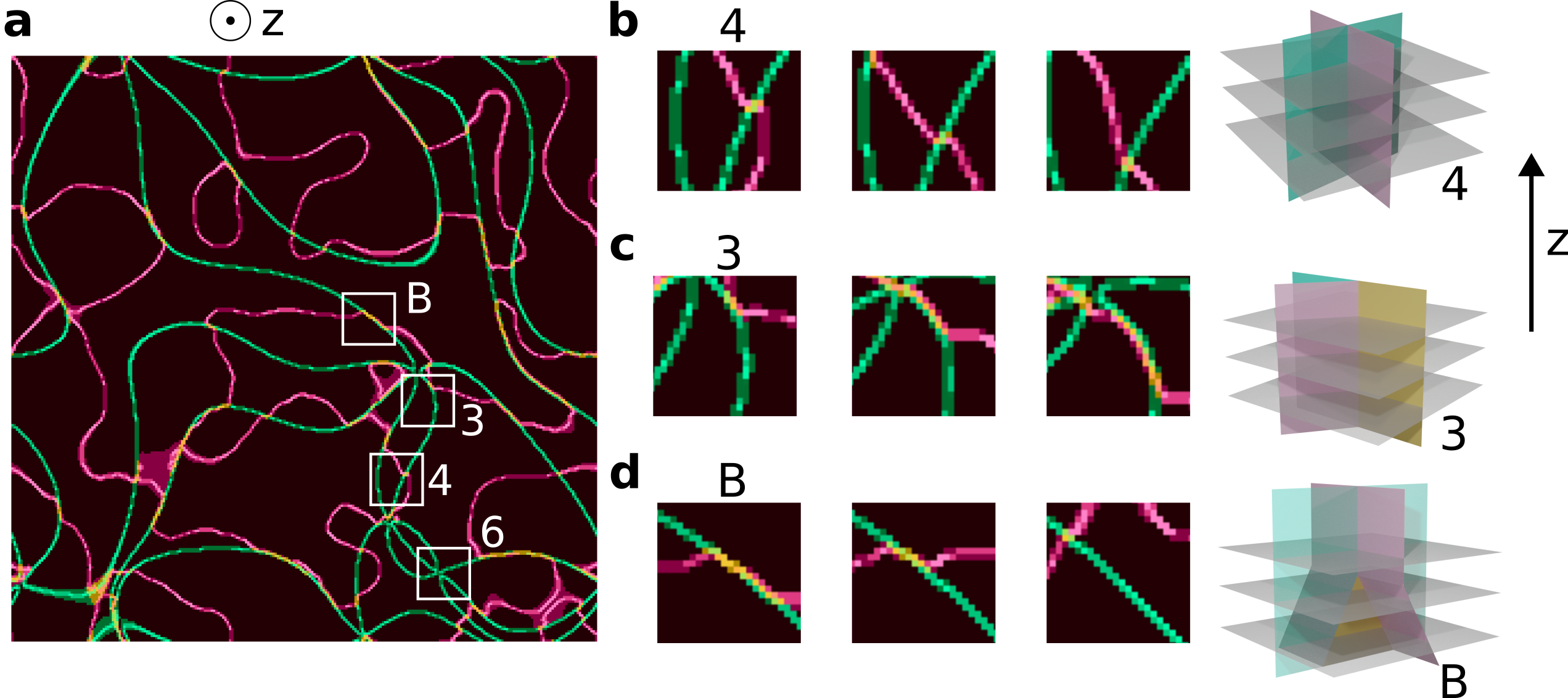}
    \caption{$\mid$ \textbf{Visualization of topological features in the simulated domain pattern.} \textbf{a} Two-dimensional slice perpendicular to the $z$-direction cutting through six-fold vortex lines, four-fold pseudo-vortex lines, three-fold pseudo-vortex lines, and bifurcations, labeled with `6', `4', `3', and `B', respectively. \textbf{b-d} Visualization of a four-fold pseudo-vortex line, a three-fold pesudo-vortex line, and a vortex line bifurcation with stacked z-cuts as schematically visualized on the right.}

    \label{fig:Data_VTK_2D_Slices}
\end{figure}

Having established the six-fold vortex lines in the data, we now continue by discussing the pseudo-vortex lines. Figure \ref{fig:Data_VTK_2D_Slices}a shows an overview of topological features observed in a 2D cut of the simulated ErMnO$_3$ system, with six-fold vortex lines, three-fold pseudo-vortex lines, four-fold pesudo-vortex lines, and bifurcations, respectively. Figures\ \ref{fig:Data_VTK_2D_Slices}b and \ref{fig:Data_VTK_2D_Slices}c then show how four-fold and three-fold pseudo-vortex lines extend into three dimensions by showing stacked 2D slices. In Fig.\ \ref{fig:Data_VTK_2D_Slices}d, we show how two three-fold pseudo-vortex lines combine into a four-fold pseudo-vortex line, thus forming a bifurcation point.

Inspection of the data using slices perpendicular to the $x$-, $y$-, and $z$-directions in supplemental data A-C shows that 
the ratio of six-fold vortex lines, four-fold pseudo-vortex lines, and three-fold pseudo-vortex lines is given by 1:7:14.
Note that these ratios depend on the parameters in Eq.\ \ref{eq:Coupling} and the domain sizes. 
These were chosen to obtain all the topological features in the limited simulation space rather than quantitatively emulating experimental data, which is also expressed by the absence of a length scale in our figures.

As mentioned, three-fold and four-fold pseudo-vortex lines are topologically protected because domain walls are closed surfaces in bulk systems. Therefore, pseudo-vortex lines must either form loops or end in a bifurcation point. Inspection of the data shows that such bifurcation points indeed appear in the simulated data, and that they are an essential ingredient
of the magnetic domain structure in three dimensions.

\subsection{Attraction between vortices and $\Psi - \Phi$ walls}

\begin{figure}[ht!]
    \centering
    \includegraphics[width=0.95\linewidth]{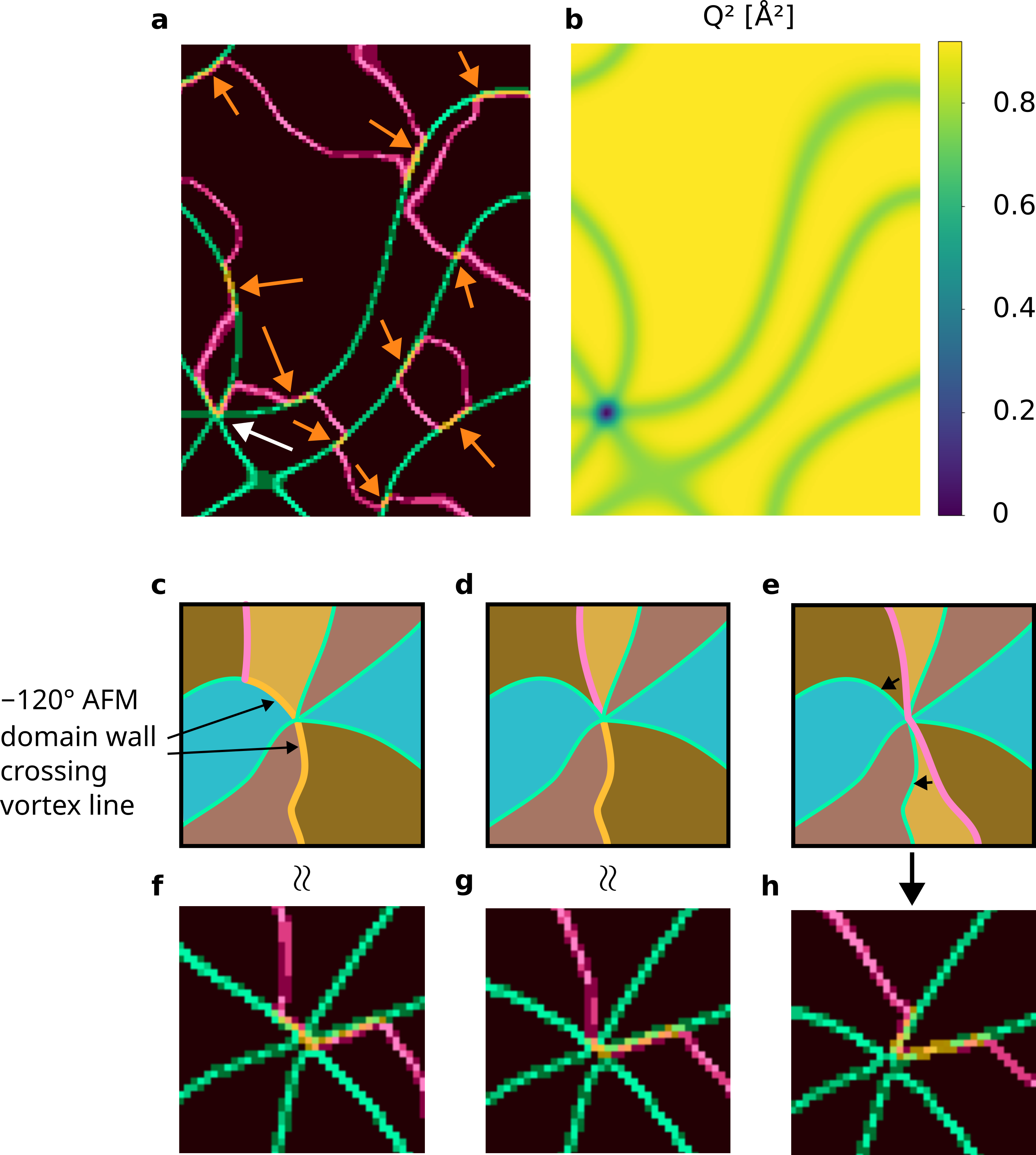}
    \caption{$\mid$ \textbf{Attraction of magnetic domain walls.} \textbf{a} Two-dimensional slice illustrating how 180° and 60° domain walls attract each other to form three-fold pseudo-vortex lines and $-120$° domain walls. A six-fold vortex with two 120° domain walls is marked with a white arrow. Orange arrows show the numerous sections where a 180° magnetic domain wall attaches to a 60° domain wall. \textbf{b} Value of the squared structural tilt amplitude $Q^2$ in the same excerpt as in (a). The squared amplitude $Q^2$ is lowered at domain walls, and even further lowered at vortex line cores. \textbf{c} Six-fold vortex with two $-120$° antiferromagnetic domain walls. \textbf{d} `Seven-fold vortex' with three-fold pseudo-vortex line and six-fold vortex-line intersecting, resulting in a 180° domain walls and a $-120$° domain wall meeting at the vortex line. \textbf{e} Sketch of a hypothetical `eight-fold vortex line', with a 180° magnetic domain wall crossing the vortex line. Black arrows illustrate the attraction of the 180° magnetic domain walls to the 60° structural domain walls. \textbf{f-h} Seven-fold vortex as a singular point in the stacked set of simulated data. }
    \label{fig:Attachment_Example}
\end{figure}

Further examining the simulated domain pattern of the system, 180° magnetic domain walls appear to be attracted to 60° magnetic domain walls with a preference for forming $-120$° magnetic domain walls in the process. Figure \ref{fig:Attachment_Example}a shows an excerpt of a 2D cut which includes numerous examples of such an attachment, highlighted by orange arrows. Furthermore, Fig.\ \ref{fig:Attachment_Example}a shows an example of a six-fold magnetic vortex line including two $-120$° domain walls rather than the usual 60° domain walls, corresponding to a $\Psi - \Phi$ domain wall crossing the vortex. It is marked with a white arrow in Fig.\ \ref{fig:Attachment_Example}a and schematically shown in Fig.\ \ref{fig:Attachment_Example}c. This remarkable feature appears multiple times in the simulated domain pattern, which can be verified by examining the 2D slice data in supplemental data A-C. 

In fact, the attraction between magnetic 60° and 180° walls with the resulting formation of $-120$° walls is related to a coupling between the magnetic order and the amplitude $Q$ of the structural order in Eq.\ \ref{eq:Coupling}. 
Inside 180° domain walls, the factor $\text{cos}^2 (\Psi - \Phi)$ of the term $A \, Q^2 \, \text{cos}^2 (\Psi - \Phi)$ does not reach the bulk minimum value. It is known from previous work and further emphasized by Fig.\ \ref{fig:Attachment_Example}b that at structural 60° $\Phi$ domain walls, the MnO$_5$ bipyramid tilt, and hence the value of $Q^2$, are lowered in comparison to the interior of the domains \cite{artyukhin_landau_2014}. Therefore, 180° domain walls in $\Psi - \Phi$ have lower energy if they coincide with a 60° structural $\Phi$ domain wall. This gives rise to the observed attraction and the associated formation of three-fold pseudo-vortex lines.
In addition to the attraction between 180° and 60° domain walls, there is also an attraction between 180° walls and six-fold vortex lines. 
Indeed, as shown in Fig.\ \ref{fig:Attachment_Example}b, $Q^2$ is even lower at vortex-line cores compared to domain walls, and therefore $\Psi - \Phi$ domain walls must be strongly attracted by structural vortex lines. %

Both types of magnetic six-fold vortex lines, that is, the common variety with six $60$° $\Psi$ domain walls and the one with two $-120$° and four $60$° $\Psi$ domain walls, coincide with structural six-fold vortex lines. 
Hence, we have a true one-to-one correspondence between structural and magnetic six-fold vortex lines, as introduced in Sec.\ \ref{sec:EquationDiscussion}.

Because 180° domain walls in $\Psi - \Phi$ are strongly attracted to structural vortex lines, one can also predict seven-fold vortex lines, where a 180° domain wall and a $-120$° vortex wall meet in at a vortex line, sketched in Fig.\ \ref{fig:Attachment_Example}d. Figures \ref{fig:Attachment_Example}f-h show that such a feature indeed appears in the simulated data. However, it does not extend into three dimensions as a line, but is rather a singular, zero-dimensional intersection point of a six-fold vortex line and a three-fold pseudo-vortex line. Such a feature will also exclusively appear in a 3D system. 

Having introduced a singular seven-fold vortex, we could also predict an eight-fold vortex point or line, as illustrated in Fig.\ \ref{fig:Attachment_Example}e. However, in the simulated data, these features are never observed. Its apparent instability is caused by immediate decay into a six-fold vortex line with two $-120$° domain walls as in Fig.\  \ref{fig:Attachment_Example}c because, as we discussed, the 180° domain walls in $\Psi - \Phi$ are attracted by neighboring 60° domain walls and would thus combine to form $-120$° domain walls. This is illustrated with the black arrows in Fig.\ \ref{fig:Attachment_Example}e.

\section{Conclusion}

In summary, we have shown that in order to gain a true understanding of emergent topological features in the multiferroic domain pattern of hexagonal manganites, a fully 3D investigation is indispensable, however experimentally and computationally complex it may be. Topological phenomena that only occur in three dimensions are essential to the understanding of the system, following from coupling processes between the distortive-ferroelectric structural order and the antiferromagnetic order. Specifically, we find a distribution of three different types of magnetic domain walls, where 60° and 180° walls attract each other to form \mbox{$-120$°} domain walls because of a suppression of the distortive-ferroelectric order at the domain walls. The three types of magnetic domain walls intersect to form six-fold vortex lines and three- and four-fold pesudo-vortex lines. In addition, we find unusual seven-fold vortex and bifurcation points, where the three-dimensional nature of the system and the attraction between magnetic 60° and 180° walls manifest most clearly.

Despite the focus on the hexagonal manganites, our results reach beyond this central class of compounds and illustrate an essential quality of multiferroic materials. 
We have shown that multiferroic domain walls exhibit properties that fundamentally exceed those of purely magnetic or electrical walls. In view of the current focus on domain-wall phenomena, multiferroic domain walls therefore deserve specific attention. Above all, we make it clear with our work that for the description of the 3D multiferroic order it is insufficient to simply extrapolate the structures seen on 2D surfaces or cross-sections into three dimensions. Novel fundamental and potentially functional phenomena whose occurrence is inherent to the third dimension will be missed.

\backmatter

\bmhead{Acknowledgements}
The authors acknowledge fruitful discussions with Sergey Artyukhin and Morgan Trassin. This work was funded by the Swiss National Science Foundation (SNSF) through Grants No. 200021\_178825 and No. 200021\_215423.

\bmhead{Competing interests}
The authors declare no competing interests.

\bmhead{Author contributions}
The project was conceived by A.M.M. Simulations were conducted by A.M.M. and L.H. Data analysis was conducted by A.M.M., with contributions from L.H. All authors contributed to the manuscript.

\bmhead{Materials \& Correspondence}
Correspondence and requests for materials should be addressed to A.M.M. or T.L.

\bmhead{Data availability} The data that support the findings of this study are available from the corresponding author upon reasonable request. In addition, a visualization of the data is available in the supplemental material.

\bmhead{Code availability} The code supporting the findings of this study is available from the corresponding author upon reasonable request.

\begin{figure}[ht!]
    \centering
    \includegraphics[width=1.\linewidth]{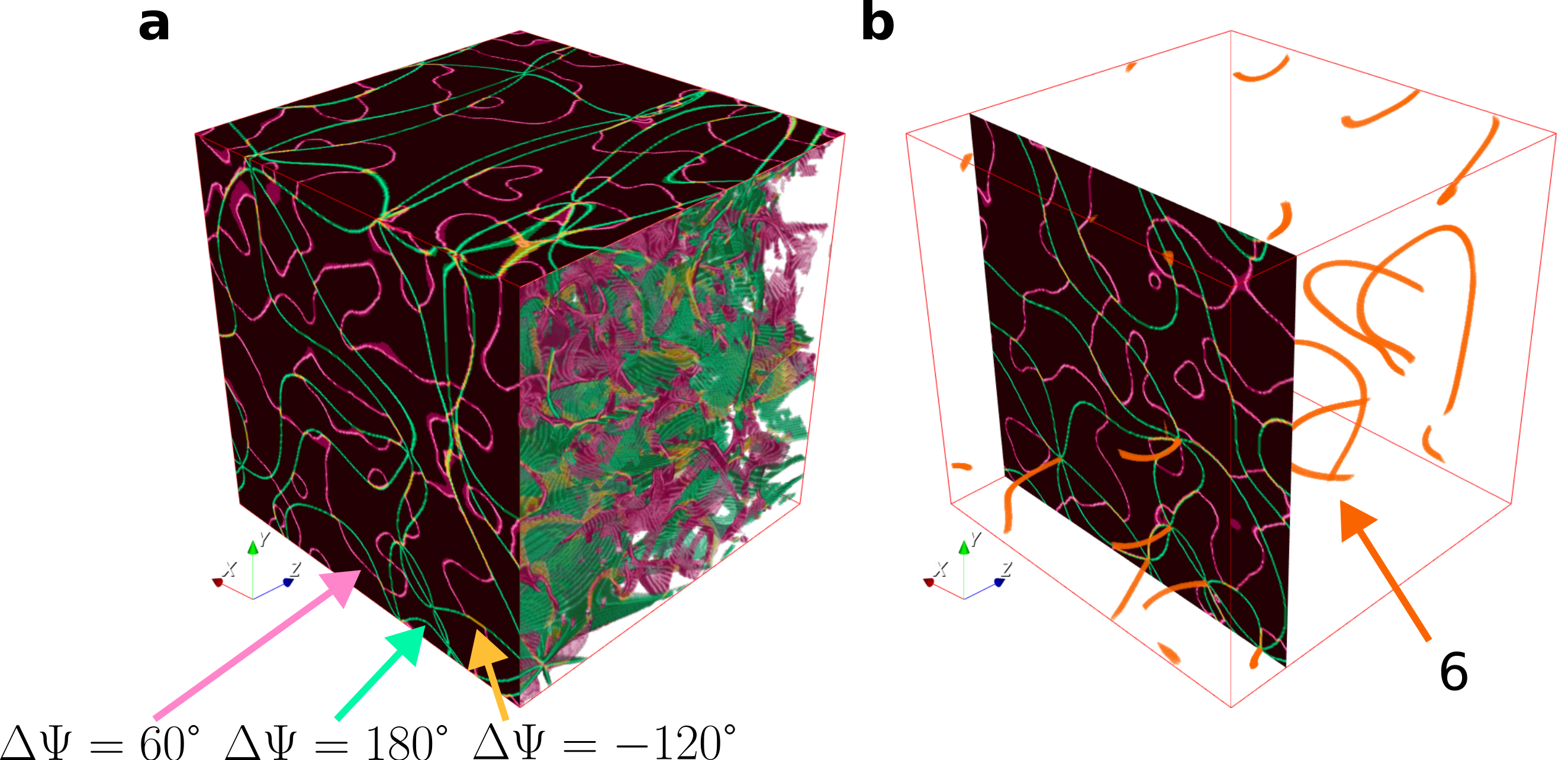}
    \caption{$\mid$ \textbf{Three-dimensional visualization of the simulated magnetic domain pattern.} Both figures visualize the same data. \textbf{a} Overview of the simulated data, with 2D cuts perpendicular to the $y$- and $z$-directions, and 3D visualization elsewhere. \textbf{b} Visualization of six-fold vortex lines (marked with `6') in the magnetic order along with an exemplary $z$-cut.
    Vortex lines always form loops which in this visualization are interrupted at the periodic boundaries of the system.}
    \label{fig:Data_VTK_3D}
\end{figure}

\bmhead{Supplementary information}
\begin{itemize}
    \item Supplemental data A-C: 2D slices through the data presented in Fig.\ \ref{fig:Data_VTK_3D} perpendicular to the x- (A), y- (B) and z-direction (C).
    \item Supplemental movie 1: 3D animation of consecutive slices presented in supplemental data C.
    \item Supplemental movie 2: 2D animation of consecutive slices presented in suppemental data C.
    \item Supplemental movie 3: 3D animation of consecutive slices presented in supplemental data C with six-fold vortex-lines visualized in orange.
\end{itemize}

\newpage

\bibliography{bibliography}%

\clearpage

\clearpage

\clearpage

\clearpage

\clearpage

\clearpage

\clearpage

\clearpage

\begin{appendices}

\section{Detailed numerical methods}
\label{app:NumericalMethods}

\textit{Parameters of the magnetic Landau expansion.} 
To simulate the magnetic order, we use the simplified Landau expansion \cite{giraldo_magnetoelectric_2021},

\begin{equation}
    F_\Psi = s_\Psi ( \nabla \Psi)^2 + A Q^2 \text{cos}^2(\Psi - \Phi).
\end{equation}

The parameters of this expansion, as taken from \cite{giraldo_magnetoelectric_2021}, are $s_\Psi = 742 \, \text{meV \AA}^2$ and $A = 2.13 \, \text{meV}$. In the phase-field simulations, an increased value of $A' = A \cdot 250$ has been used. This adaption, which was also utilized in \cite{giraldo_magnetoelectric_2021}, reduces the width of the magnetic domain walls to ensure that these become similar in thickness to the domain wall of the structural order. This approximation does not affect the topology of the system or its consequences, and it greatly improves the computational efficiency of the phase-field simulations.

\textit{Computational details.}
For simulations of both the structural and the antiferromagnetic order we use a regular computational lattice of $256 \times 256 \times 256$ with lattice spacing $\Delta x = \Delta y = \Delta z = 1 \text{\AA}$. For the simulations of the ferroelectric domain pattern, we initialize $Q_x = Q \cdot \, \text{cos} \Phi$, $Q_y = Q \cdot \, \text{sin} \Phi $, and $\mathcal{P}$ with a random, uniform distribution of the interval \mbox{[$-$0.1\,\AA, 0.1\,\AA]}. 
The dynamics of the system is then governed by the Ginzburg-Landau equation

\begin{equation}
    \frac{\partial \eta}{\partial t} = - \frac{\delta f}{ \delta \eta},
    \label{eq:Landau_Derivative_App}
\end{equation}

which we solve using a finite-difference Runge-Kutta 4 solver. We use a time step of $\Delta t = 5 \cdot 10^{-3}$ and perform $N = 3 \cdot 10^4$ iteration steps. We choose periodic boundary conditions to simulate bulk behavior. The ferroelectric order of the system is then frozen for the reasons detailed in the main text, and the distribution of values of $Q$ and $\Phi$ are used for the simulations of the antiferromagnetic order.

We initialize the antiferromagnetic order with random, uniform values of $\Psi$ in the interval $(-\pi, \pi]$. The evolution of the system is then again governed by Eq.\ \ref{eq:Landau_Derivative_App}, which we solve with the same finite-difference Runge-Kutta 4 solver as for the ferroelectric order. We choose a time step of $\Delta t = 2 \cdot 10^{-3}$ and perform $N = 1.4 \cdot 10^5$ iteration steps. Again, we choose periodic boundary conditions to simulate a bulk crystal.

\textit{Data Analysis.}
To obtain all three types of magnetic domain walls, we first compute $\mathcal{L} = \text{sin}(3 \Psi)$, a 1D field with nonzero gradients at 60° and 180° magnetic domain walls.
We then compute $\mathcal{P} \mathcal{L}$, a field with nonzero gradients at 180° and $-120$° domain walls \cite{giraldo_magnetoelectric_2021}. Furthermore, we know from Eq.\ \ref{eq:Condition} that the domain walls of the $\mathcal{P}$ field correspond to magnetic 60° and $-120$° domain walls. 
By computing gradients of all three fields, the 180°, 60° and $-120$° magnetic domain walls can be extracted.
Vortex-line and pseudo-vortex line lengths are estimated choosing test planes perpendicular to the $x$-, $y$-, and $z$-directions and counting the number of line intersections with these test planes \cite{underwood_quantitative_1973}. %

\end{appendices}

\end{document}